\documentclass[11pt]{article}

\usepackage{amssymb,latexsym,amsmath}
\usepackage{xcolor}
\usepackage{bm}

\usepackage{hyperref}

\begin{document}

\newtheorem{theorem}{Theorem}[section]

\newtheorem{proposition}[theorem]{Proposition}

\newtheorem{lemma}[theorem]{Lemma}

\newtheorem{corollary}[theorem]{Corollary}

\newtheorem{definition}[theorem]{Definition}

\newtheorem{remark}[theorem]{Remark}

\newtheorem{exempl}{Example}[section]

\newenvironment{exemplu}{\begin{exempl}  \em}{\hfill $\surd$

\end{exempl}}

\newcommand{\ea}{\mbox{{\bf a}}}
\newcommand{\eu}{\mbox{{\bf u}}}
\newcommand{\ep}{\mbox{{\bf p}}}
\newcommand{\ed}{\mbox{{\bf d}}}
\newcommand{\eD}{\mbox{{\bf D}}}
\newcommand{\eK}{\mathbb{K}}
\newcommand{\eL}{\mathbb{L}}
\newcommand{\eB}{\mathbb{B}}
\newcommand{\ueu}{\underline{\eu}}
\newcommand{\ueo}{\overline{u}}
\newcommand{\oeu}{\overline{\eu}}
\newcommand{\ew}{\mbox{{\bf w}}}
\newcommand{\ef}{\mbox{{\bf f}}}
\newcommand{\eF}{\mbox{{\bf F}}}
\newcommand{\eC}{\mbox{{\bf C}}}
\newcommand{\en}{\mbox{{\bf n}}}
\newcommand{\eT}{\mbox{{\bf T}}}
\newcommand{\eV}{\mbox{{\bf V}}}
\newcommand{\eU}{\mbox{{\bf U}}}
\newcommand{\ev}{\mbox{{\bf v}}}
\newcommand{\eve}{\mbox{{\bf e}}}
\newcommand{\uev}{\underline{\ev}}
\newcommand{\eY}{\mbox{{\bf Y}}}
\newcommand{\eP}{\mbox{{\bf P}}}
\newcommand{\eS}{\mbox{{\bf S}}}
\newcommand{\eJ}{\mbox{{\bf J}}}
\newcommand{\leb}{{\cal L}^{n}}
\newcommand{\eI}{{\cal I}}
\newcommand{\eE}{{\cal E}}
\newcommand{\hen}{{\cal H}^{n-1}}
\newcommand{\eBV}{\mbox{{\bf BV}}}
\newcommand{\eA}{\mbox{{\bf A}}}
\newcommand{\eSBV}{\mbox{{\bf SBV}}}
\newcommand{\eBD}{\mbox{{\bf BD}}}
\newcommand{\eSBD}{\mbox{{\bf SBD}}}
\newcommand{\ecs}{\mbox{{\bf X}}}
\newcommand{\eg}{\mbox{{\bf g}}}
\newcommand{\paromega}{\partial \Omega}
\newcommand{\gau}{\Gamma_{u}}
\newcommand{\gaf}{\Gamma_{f}}
\newcommand{\sig}{{\bf \sigma}}
\newcommand{\gac}{\Gamma_{\mbox{{\bf c}}}}
\newcommand{\deu}{\dot{\eu}}
\newcommand{\dueu}{\underline{\deu}}
\newcommand{\dev}{\dot{\ev}}
\newcommand{\duev}{\underline{\dev}}
\newcommand{\weak}{\rightharpoonup}
\newcommand{\weakdown}{\rightharpoondown}
\renewcommand{\contentsname}{ }

\title{A stochastic version and a Liouville theorem for hamiltonian inclusions with convex dissipation}
\author{\href{http://imar.ro/~mbuliga/}{Marius Buliga} \\ Institute of Mathematics of the Romanian Academy}
\date{This version: 27.07.2018\footnote{to Matei, for his anniversary}}


\maketitle

\begin{abstract}
The statistical counterpart  of the formalism of hamiltonian systems with convex dissipation \cite{bham} \cite{MBGDS1} is a completely open subject. Here are described a  stochastic version of the SBEN principle and  a Liouville type theorem which uses  a minimal dissipation cost functional.
\end{abstract}

\section{Introduction}

In definition \ref{sbenprob} is proposed the stochastic equation in $z(t) = (q,p)(t)$
$$
(\dot{q}, \dot{p})(t) - XH(t,(q,p)(t)) \, = \, \dot{z}_{D}(t) 
$$
where $\displaystyle \dot{z}_{D}(t)$ is taken randomly with probability
$$
\pi(t,q,p, \dot{z}_{D}) \, = \, \frac{exp\left(- \beta \left[\phi \left(\dot{z}_{D} + XH(t,z)\right) + \phi^{*\omega} \left( \dot{z}_{D}\right) + \omega \left( XH(t,q,p), \dot{z}_{D} \right) \right]\right)}{Z(t,z)}
$$
$XH(t,q,p)$ is the symplectic gradient of the hamiltonian $H(t,q,p)$ and $\phi$ is a convex dissipation potential, with symplectic Fenchel conjugate $\displaystyle \phi^{*\omega}$, definition \ref{dssubpolar}. 

In the limit of the parameter $\beta>0$ to $+\infty$ we obtain a hamiltonian inclusion with convex dissipation \cite{bham}. 
$$
\dot{z}(t) \, - \, XH(t,z(t)) \, \in \partial^{\omega} \phi(\dot{z}(t))
$$
The evolution is not symplectic in general, but  theorem \ref{gibbslikethm} gives a Liouville type theorem based on the minimization of a dissipation cost.

This article is motivated by the paper Oueslati,  Nguyen and de Saxc\' e \cite{conf}. There, the authors start by presenting the formalism of hamiltonian systems with convex dissipation from Buliga \cite{bham}, in the more evolved form of the so called symplectic Brezis-Ekeland-Nayroles (SBEN) principle from Buliga and de Saxc\'e \cite{MBGDS1}. Then, they continue by two very interesting suggestions, or claims. These are (with the notations used in the present paper): 
\begin{enumerate}
\item[(a)] (\cite{conf} section 6) that a more general formalism can be constructed, by replacing the dissipation term $\displaystyle \phi(\dot{z}(t)) + \phi^{*\omega}( \dot{z}_{D}(t) )$ with a "symplectic bipotential" $\displaystyle b\left(\dot{z}(t), \dot{z}_{D}(t) \right) $. Bipotentials, introduced \cite{gds2} and applied by de Saxc\'e to a number of problems in soil mechanics,  plasticity, damage or friction, see the review \cite{gds1},  are an extension of convex analysis which can cover non associated constitutive laws. A rigorous theory has been constructed in a series of papers by the author together with de Saxc\' e and Vall\' ee, see for example the review \cite{gds3}.  
\item[(b)] (\cite{conf} section 7) that the dissipative (according to the model) transition between macrostates (i.e.  Gibbs measures) is an optimal transportation problem, in the sense that it minimizes a cost functional which appears naturally in the symplectic BEN principle. 
\end{enumerate}

The suggestion (a), although straightforward, needs to be taken seriously by passing the symplectic bipotential replacement through the stages of the construction presented in \cite{MBGDS1}. Examples obtained from the adaptation of well known bipotentials to the symplectic case may also be revealing. 

The claim (b) is lightly sketched in \cite{conf}, with no proofs. This claim motivates the present article. Thanks to the fact that de Saxc\' e contacted me after the acceptance of the paper \cite{conf}. but before publication, I became interested into the subject and proposed a rigorous Liouville type theorem which I think it is what the authors really wanted to say in claim (b). The problem of a statistical treatment of Hamiltonian systems with convex dissipation is completely open at the moment and (a correct version of) claim (b) seems to be a good start towards such a theory.

The subject of extension of the hamiltonian formalism for dissipative systems is vast. Further is a non-exhaustive list of such works. Aubin \cite{aubin2}, Aubin, Cellina and Nohel \cite{aubin}, Rockafellar \cite{rocka} proposed several ways to extend the hamiltonian and lagrangian mechanics. In Bloch, Krishnaprasad, Marsden and Ratiu \cite{bloch} are explored hamiltonian systems with a Rayleigh dissipation. In several papers Mielke and collaborators built a theory of quasistatic rate-independent systems:  Mielke and Theil \cite{mielketh99}, Mielke \cite{mielke}, and presented many applications, among them Mielke and Roub\'{\i}\v{c}ek \cite{MR06b}. See also  Visintin \cite{visintin}.

\section{Hamiltonian inclusions with convex dissipation}

\paragraph{Notations.} 
Let $\displaystyle X$ be a real topological vector space of states $q \in X$ and $\displaystyle Y$ a dual space of momenta $\displaystyle p \in Y$. The duality is denoted by 
$$ (q,p) \in X \times Y \, \mapsto \langle q , p \rangle \in \mathbb{R}$$
with the following, usual properties: is bilinear, continuous and for any linear and continuous functions $\displaystyle L: X \rightarrow \mathbb{R}$, $\displaystyle G: Y \rightarrow \mathbb{R}$ there exist $q \in X$, $\displaystyle p \in Y$ such that $\displaystyle L(\cdot) = \langle \cdot, p \rangle$ and $\displaystyle G(\cdot) = \langle q, \cdot \rangle$. 

Denote by $\displaystyle N = X \times Y$. This space is the dual of $\displaystyle N^{*} = Y \times X$, by the duality: 
$$ \langle \langle (p_{1}, q_{1}) , (q_{2}, p_{2}) \rangle\rangle \, = \, \langle q_{1}, p_{2} \rangle \, + \, \langle q_{2}, p_{1} \rangle
$$
We shall use the functions $\displaystyle J: N \rightarrow N^{*}$, $\displaystyle J^{*}: N^{*} \rightarrow N$, 
$$ J(q,p) \, = \, (-p, q) \, , \, J^{*}(p,q) \, = \, (-q,p)$$
Clearly $\displaystyle -J^{*}J$ is the identity of $\displaystyle N$ and $\displaystyle -J J^{*}$ is the identity of $\displaystyle N^{*}$. 

The space $N$ is symplectic, with the symplectic form: for any $\displaystyle z_{1} = (q_{1}, p_{1})$ , $\displaystyle z_{2} = (q_{2}, p_{2})$ 
$$\omega(z_{1}, z_{2}) \, = \, \langle \langle J z_{1}, z_{2} \rangle \rangle \, = \, \langle q_{1}, p_{2} \rangle \, - \, \langle q_{2} , p_{1} \rangle $$

For any differentiable function $\displaystyle H: N \rightarrow \mathbb{R}$ the gradient of $H$ at a point $z \in N$ is an element $\displaystyle DH(z) \in N^{*}$ and the symplectic gradient of $H$ is $XH(z) \in N$ defined by 
$$ XH(z) \, = \, - J^{*} \, DH(z)
$$

\begin{definition} For any convex, lower semi-continuous function $$F: X \times Y \rightarrow \mathbb{R}\cup \left\{+\infty\right\}$$ the symplectic sub-differential of $F$ at a point $z = (x,y) \in X \times Y$ such that $F(z) < + \infty$  is the set 
$$\displaystyle \partial^{\omega} F (z) \, = \, \left\{ 
z' \in X \times Y \mbox{ : } \forall \, z" \in X \times Y \quad  
F(z+z") \,  \geq \,  F(z) \, + \, 
\omega (z' , z") \right\} $$ 
The symplectic Fenchel transform (named also the symplectic polar) of $F$ is the function: 
$$F^{*\omega}(z') \, = \, \sup \left\{ \omega(z',z) - F(z) \mbox{ : } z \in X \times Y \right\}$$
\label{dssubpolar}
\end{definition}

 The relations between the usual sub-differential and polar and their symplectic versions are the following. 

\begin{proposition}
With the notations from Definition \ref{dssubpolar} we have: 
\begin{enumerate}
\item[-] $\displaystyle z' \, \in \, \partial^{\omega} F(z)$ is equivalent with $\displaystyle J z' \, \in \, \partial F(z)$
\item[-] $\displaystyle - J^{*} z' \, \in \, \partial^{\omega} F(z)$ is equivalent with $\displaystyle z' \, \in \, \partial F(z)$
\item[-] $\displaystyle F^{* \omega}(z) \, = \, F^{*}(Jz')$
\item[-] $\displaystyle F^{* \omega}(J^{*} z) \, = \, F^{*}(-z')$
\item[-] the symplectic Fenchel inequality: for any $z,z' \in X \times Y$ we have 
$$F(z)\, + \, F^{*\omega} (z')  \, \geq \, \omega(z',z)$$ 
and the equality is attained if and only if $\displaystyle z' \in \partial^{\omega} F (z)$. 
\end{enumerate}
\label{ususymp}
\end{proposition}

\paragraph{The main equation.} 
The following are given: 
\begin{enumerate}
\item[(H)] a smooth hamiltonian $H = H(t,x,y) = H(t,z)$, 
\item[(D)] a convex lower semicontinuous dissipation potential $\displaystyle \phi= \phi( \dot{z})$,  
$$\phi:  X \times Y \rightarrow \mathbb{R} \cup \left\{ + \infty \right\}$$  
\end{enumerate}
In \cite{MBGDS1}  remark 4.4 is explained that for physical reasons it is convenient to make the following hypothesis:   for any $z, z'\in X\times Y $
\begin{equation}
 \phi(z) + \phi^{*\omega}(z') \geq 0  
\label{D}
\end{equation}
This hypothesis is not used here, until corollary \ref{cworkpump}.

 The following equation has first appeared as a "hamiltonian inclusion with convex dissipation" \cite{bham}. 
\begin{equation}
\dot{z}(t) \, - \, XH(t,z(t)) \, \in \partial^{\omega} \phi(\dot{z}(t))
\label{hamdis}
\end{equation}
Any solution of this equation is a curve $z: [0,T] \rightarrow X \times Y$.

This equation can be seen as a dynamical version of  Mielke \cite{mielke} notion of evolution in rate-independent systems.

 A rewrite of this equation in \cite{MBGDS1}  is called the symplectic BEN (or SBEN) principle because it is shown there that in the quasistatic approximation, for well chosen hamiltonian and dissipation potential, there are recovered the variational principles of Brezis-Ekeland \cite{Brezis Ekeland 1976} and Nayroles \cite{Nayroles 1976}.

\begin{definition} (The symplectic BEN principle.) An evolution curve  $t \in [0,T] \mapsto z(t) \in X \times Y$ satisfies the SBEN (symplectic Brezis-Ekeland-Nayroles) principle for the hamiltonian $H$ and dissipation potential $\phi$ if for almost any $t \in [0,T]$ we have a decomposition of $\displaystyle \dot{z}$  into "conservative"  and "dissipative" parts:
\begin{equation}
\dot{z} = \dot{z}_{C} + \dot{z}_{D} \quad , \quad   \dot{z}_{C} = X H (z) \, \, , \, \, \dot{z}_{D} = \dot{z} - XH(z)   
\label{decompo}
\end{equation}
such that 
\begin{equation}
\phi(\dot{z}(t)) + \phi^{*\omega}( \dot{z}_{D}(t) ) = \omega(\dot{z}_{D}(t),\dot{z}(t))   
\label{sben1}
\end{equation}
\label{dsben1}
\end{definition}

The next proposition (\cite{MBGDS1} proposition 4.3) shows (a) that indeed the SBEN principle is equivalent with (\ref{hamdis}) and (b) that it is equivalent with a variational principle based on the minimization of the functional (\ref{sben3}). In the very particular case of application to elastoplasticity and quasistatic evolution, this variational principle becomes the one of Brezis-Ekeland and Nayroles. 

\begin{proposition}
 An evolution curve  $t \in [0,T] \mapsto z(t) \in X \times Y$ satisfies the SBEN principle for the hamiltonian $H$ and dissipation potential $\phi$ if and only if it satisfies one of the following: 
\begin{enumerate}
\item[(a)] for almost  every $t \in [0,T]$ 
\begin{equation}
\dot{z}(t) - X H(t,z(t) ) \, \in \, \partial^{\omega} \phi(\dot{z})   
\label{sben3}
\end{equation}
\item[(b)] the evolution curve minimizes the functional 
\begin{equation}
\Pi(z') = \int_{0}^{T} \left\{ \phi(\dot{z'}(t)) + \phi^{*\omega}(\dot{z}'_{D}(t)) -  \frac{\partial H}{\partial t}(t,z'(t)) \right\} \mbox{ dt} \,  + 
\label{sben2}
\end{equation}
$$ + \,  H(T, z'(T)) $$
among all curves $z': [0,T] \rightarrow X \times Y$ such that $\displaystyle z'(0) = z(0)$. 
\end{enumerate}
\label{pequisben}
\end{proposition}

\paragraph{A stochastic version.} Since by the Fenchel inequality we have
\begin{equation}
\phi(\dot{z}(t)) + \phi^{*\omega}(\dot{z}') -  \omega \left( \dot{z}', \dot{z} \right) \, \geq \, 0
\label{fenchel}
\end{equation}
 by (\ref{decompo}) and (\ref{sben1}) the vector $\displaystyle \dot{z}' \, = \, \dot{z} - XH(z)$ minimizes the left hand side of the Fenchel inequality and the minimum is equal to $0$. This strongly suggest the following definition.

\begin{definition}
(Stochastic SBEN) Let $\beta>0$ be an inverse temperature parameter. We introduce for any $z = (q,p)$  a probability on the space of velocities 
$$\displaystyle \dot{z}_{D} = \left( \dot{q}_{D} ,  \dot{p}_{D} \right)$$
 with  density 
\begin{equation}
\pi(t,z, \dot{z}_{D}) \, = \, \frac{exp\left(- \beta \left[\phi \left(\dot{z}_{D} + XH(t,z)\right) + \phi^{*\omega} \left( \dot{z}_{D}\right) + \omega \left( XH(t,z), \dot{z}_{D} \right) \right]\right)}{Z(t,z)}
\label{pdens}
\end{equation} 
We define a random evolution 
\begin{equation}
\dot{z}(t) - XH(t,z(t)) \, = \, \dot{z}_{D}(t) 
\label{bigrandomevo}
\end{equation}
 such that $\displaystyle \dot{z}_{D}(t)$ is picked with probability $\displaystyle \pi(t,z(t), \cdot)$. 
\label{sbenprob}
\end{definition}
In the limit of the inverse temperature parameter $\beta$ to $+ \infty$ we retrieve the evolution according to the SBEN principle. 

It is instructive to detail the case when 
$$H(t,q,p) \, = \, \frac{1}{2m} \| p \|^{2} + V(q) \, \, \, , \, \, \, \phi(\dot{z}) \, = \, \Phi(\dot{q})$$
By direct computation using definition \ref{dssubpolar} and proposition \ref{ususymp} we have 
$$\phi^{*\omega}(\dot{z}_{D}) \, = \,  \Phi^{*}\left( - \dot{p}_{D} \right) $$
if and only if $\displaystyle \dot{q}_{D} = 0$, i.e. 
$$ \dot{q}_{D} \, = \, \dot{q} -  D_{p} H(q,p) \, = \, \dot{q} -  \frac{1}{m} p \, = \, 0$$ 
We introduce the random force $$\displaystyle \eta \, = \, \dot{p}_{D}$$
which gives the evolution equation (\ref{bigrandomevo}) with the form
\begin{equation}
m \ddot{q} + D_{q} V(q) \, = \, \eta
\label{randomevo}
\end{equation}
where the random force is picked according to the probability density
\begin{equation}
\pi(q,\dot{q}, \eta) \, = \, \frac{exp\left(- \beta \left[ \Phi^{*} \left( - \eta  \right) + \langle \eta, \dot{q} \rangle  \right]\right)}{Z'(q,\dot{q})}
\label{pieta}
\end{equation}
Indeed: 
$$ \phi \left(\dot{z} \right) + \phi^{*\omega} \left( \dot{z} - XH(t,z)\right) + \omega \left( XH(t,z), \dot{z} \right) \, = \, $$ 
$$ \, = \, \Phi(\dot{q}) + \Phi^{*}(-\eta) + \omega \left( XH(q,p), (\dot{q}, \dot{p}) \right) \, = \, $$ 
$$ \, = \,  \Phi(\frac{1}{m} p) + \Phi^{*}(-\eta) +  \langle D_{q} H(q,p), \dot{q} \rangle + \langle \dot{p}, D_{p}H(q,p) \rangle \, = \, $$ 
$$\, = \,  \Phi(\frac{1}{m} p) + \Phi^{*}(-\eta) +  \langle D_{q} V(q) + \dot{p}, \dot{q} \rangle \, = \, $$ 
$$ \, = \,  \Phi(\frac{1}{m} p) +  \Phi^{*}(-\eta)  + \langle \eta, \dot{q} \rangle $$ 
When we plug this into (\ref{pdens}) the term $\displaystyle \Phi(\frac{1}{m} p)$ disappears and we obtain the probability density (\ref{pieta}) for the random force, with a corresponding modified normalization constant $\displaystyle Z'(q,\dot{q})$.

\section{A Liouville type theorem}

In this section we suppose that the symplectic space $(X \times Y, \omega)$ is finite dimensional and endowed with a Lebesgue measure $\mbox{dz}$.

Suppose $\displaystyle \psi: [0,T] \times ( X \times Y) \rightarrow X \times Y$ is a flow such that  for all $z \in X \times Y$  $\psi(0,z)=z$ and  the curve $t \mapsto \psi(t,z)$ is an evolution curve which satisfies the SBEN principle. Contrary to the non dissipative case (i.e. $\phi=0$), there is no reason for the maps $\psi(t,\cdot)$ to be volume preserving or symplectomorphisms, or even invertible.  

Denote by $Paths(H,F)$  the collection   of flows $\displaystyle \Psi: [0,T] \times ( X \times Y) \rightarrow X \times Y$ which are smooth with respect to time and  such that  for all $z \in X \times Y$  $\Psi(0,z)=z$ and $\Psi(T,z)=F(z)$. 
For any $\Psi \in Paths(H,F)$  let: 
$$\dot{\Psi}_{D}(t,z) \, = \, \dot{\Psi}(t,z) \, - \, XH(t,\Psi(t,z)) $$
and remark that 
$$\langle\langle  DH(t,\Psi(t,z)), \dot{\Psi}(t,z) \rangle\rangle \, = \, - \omega(\dot{\Psi}_{D}(t,z), \dot{\Psi}(t,z))  $$

For any $\Psi \in Paths(H,F)$ we introduce a curve of   Gibbs measures $\displaystyle t \in [0,T] \mapsto \mu_{t}$  defined by: for any Borel set $ B \subset X \times Y$ 
$$\mu_{t}(B) \, = \, \int_{B}  exp\left[ - \left(\alpha + \beta H(t,\Psi(t,z)) \right)   \right] \mbox{ dz} \, = \, \int_{B} 1 \mbox{ d} \mu_{t}(z)  $$
We therefore have $\displaystyle \mu_{0}=\mu(H, id)$ and $\displaystyle \mu_{T}=\mu(H, F)$. 

Remark that we do not suppose that the flow $\Psi$ is volume preserving, therefore $\displaystyle \mu_{t}$ is not a transport of $\displaystyle \mu_{0}$ by the flow $\Psi$, i.e. in general 
$$\mu_{t}(B) \not = \mu_{0}(\Psi(B))  $$

Define the dissipation cost of the flow $\Psi \in Paths(H,F)$ over the set $B$: 
$$C(\Psi)(B) \, = \,  \int_{0}^{T} \int_{B} \left[ \phi(\dot{\Psi}(t,z)) + \phi^{*\omega}(\dot{\Psi}_{D}(t,z) ) \, - \,  \frac{\partial H}{\partial t}(t, \Psi(t,z)) \right]  \mbox{ d} \mu_{t}(z) 
\mbox{ dt}$$
 with $\displaystyle \mu(H, id), \mu(H, F)$ as parameters.

\begin{theorem}
Let $\displaystyle \mu(H, id), \mu(H, F)$ be two  Gibbs measures. Then for any flow $\Psi \in Paths(H,F)$ and for any Borel set $B \subset X \times Y$ we have the inequality
\begin{equation}
\mu(H, F)(B) - \mu(H, id)(B) \, \leq \,  \beta \, C(\Psi)(B)
\label{costineq}
\end{equation}
The minimal dissipation cost over $B$ is attained by flows $\Psi \in Paths(H,F)$ which for almost every $z \in B$ and $t \in [0,T]$ satisfy the SBEN principle 
$$\dot{\Psi}_{D}(t,z) \in \partial^{*\omega} \phi \left( \dot{\Psi}(t,z)\right)$$
\label{gibbslikethm}
\end{theorem}

\paragraph{Proof.}
With the notations made previously we have $\displaystyle \mu_{T} =\mu(H, F)$ and  $\displaystyle \mu_{0}=\mu(H, id)$, therefore 

$$\mu(H, F)(B) - \mu(H, id)(B) \, =  \, \mu_{T}(B) - \mu_{0}(B) \, = \, $$ 
$$= \,  \int_{0}^{T} \frac{d}{dt} \left( \int_{B}  exp\left[ - \left(\alpha + \beta H(t,\Psi(t,z)) \right)   \right] \mbox{ dz} \right) \mbox{ dt} \, = \, $$ 
$$ = \,  \left( -\beta \right)  \int_{0}^{T} \int_{B} \left[ \frac{\partial H}{\partial t}(t, \Psi(t,z)) \, + \, \langle\langle DH(t,\Psi(t,z)), \dot{\Psi}(t,z)\rangle\rangle \right]  \mbox{ d} \mu_{t}(z) \mbox{ dt} \, = \, $$
$$= \, \beta \int_{0}^{T} \int_{B} \left[ \omega(\dot{\Psi}_{D}(t,z), \dot{\Psi}(t,z)) \, - \,  \frac{\partial H}{\partial t}(t, \Psi(t,z)) \right]  \mbox{ d} \mu_{t}(z) 
\mbox{ dt} \, \leq \, $$ 
$$ \leq \, \beta \int_{0}^{T} \int_{B} \left[ \phi(\dot{\Psi}(t,z)) + \phi^{*\omega}(\dot{\Psi}_{D}(t,z) ) \, - \,  \frac{\partial H}{\partial t}(t, \Psi(t,z)) \right]  \mbox{ d} \mu_{t}(z) 
\mbox{ dt} $$
The last inequality comes from the symplectic Fenchel inequality. We recognize in the last term the dissipation cost of the flow $\Psi$ over $B$, therefore (\ref{costineq}) is proved. 

Moreover, the equality happens if and only if for almost every $z \in B$ and $t \in [0,T]$ we have: 
$$ \omega(\dot{\Psi}_{D}(t,z), \dot{\Psi}(t,z)) \, = \, \phi(\dot{\Psi}(t,z)) + \phi^{*\omega}(\dot{\Psi}_{D}(t,z) ) $$
which by Proposition \ref{pequisben} (a) is equivalent with 
$$\dot{\Psi}_{D}(t,z) \in \partial^{*\omega} \phi \left( \dot{\Psi}(t,z)\right)$$
which ends the proof. \hfill $\square$

\begin{corollary}
If the hamiltonian has the form: 
\begin{equation}
H(t, q, p) \, = \, H(q,p) - \langle f(t), q \rangle
\label{speh}
\end{equation}
and the dissipation potential satisfies (\ref{D}) then 
\begin{equation}
\mu(H, F)(B) - \mu(H, id)(B) \, \geq  \, \beta \int_{0}^{T}   \langle \frac{d}{dt}f(t), \int_{B} q(t,z)  \mbox{ d} \mu_{t}(z) \rangle   \mbox{ dt}
\label{workpump}
\end{equation}
\label{cworkpump}
\end{corollary}

\paragraph{Proof.}
We proceed as in the proof of theorem \ref{gibbslikethm}.  We choose the flow $\displaystyle \Psi(t,z)=(q(t,z),p(t,z))$ to be a solution of SBEN. Then we use  (\ref{D}) and the special form of the hamiltonian (\ref{speh}) to obtain 
$$\mu(H, F)(B) - \mu(H, id)(B) \, \geq  \, - \beta \int_{0}^{T} \int_{B}   \frac{\partial H}{\partial t}(t, \Psi(t,z))   \mbox{ d} \mu_{t}(z) \mbox{ dt}  \, = \, $$
$$= \, \beta \int_{0}^{T}   \langle \frac{d}{dt}f(t), \int_{B} q(t,z)  \mbox{ d} \mu_{t}(z) \rangle   \mbox{ dt} $$
which gives the inequality we are after. \hfill $\square$

We recognize in the right hand side of (\ref{workpump}) the average over $B$ of the work done by the external forces. The Corollary \ref{cworkpump} implies that if  the average work by the external forces is positive then $\displaystyle \mu(H, F)(B) \geq \mu(H, id)(B)$.

\end{document}